\documentstyle[aps, epsf, psfig]{revtex}

\newcommand{\pr}{\prime}
\newcommand{\na}{\nabla}

\newcommand{\ep}{\epsilon}

\newcommand{\vphi}{\varphi}

\newcommand{\lraw}{\longrightarrow}

\newcommand{\pa}{\partial}

\newcommand{\sla}[1]{\slash\!\!\! #1}
\begin{document}

\draft
\title{Running of Axial Coupling Constant}
\author{Xiao-Jun Wang\footnote{E-mail address: wangxj@mail.ustc.edu.cn}}
\address{Center for Fundamental Physics,
University of Science and Technology of China\\
Hefei, Anhui 230026, P.R. China}
\author{Mu-Lin Yan\footnote{E-mail address: mlyan@staff.ustc.edu.cn}}
\address{CCST(World Lad), P.O. Box 8730, Beijing, 100080, P.R. China \\
  and \\
Center for Fundamental Physics,
University of Science and Technology of China\\
Hefei, Anhui 230026, P.R. China\footnote{mail
address}}
\date{\today}
\maketitle

\begin{abstract}
We illustrate that both of massless pseudoscalar meson fields and 
constitutent quark fields are dynamical field degrees of freedom in energy
region between chiral symmetry spontaneously broken scale and quark
confinement scale. This physical configuration yields a renormalization
running axial coupling constant $g_{_A}(\mu)$. The one-loop
renormalization prediction $g_{_A}(\mu=m_{_N})=0.75$ and 
$g_{_A}(\mu=m_\pi)=0.96$ agree with $\beta$ decay of neutron and
$\pi^0\rightarrow\gamma\gamma$ respectively. We also calculate
$\rho^\pm\rightarrow\gamma\pi$ and $\omega\rightarrow\gamma\pi$ decays up
to the next to leading order of $N_c^{-1}$ expansion and including all
order effects of vector meson momentum expansion. The results strongly
support prediction of renormalization running, $g_{_A}(\mu=m_\rho)=0.77$.
\end{abstract}

The asymptotic freedom, chiral symmetry spontanously broken(CSSB) and
quark confinement(QC) are essential features of QCD, which are related to
the non-Abelian guage symmetry structure of QCD. They divide
QCD-description of strong interaction into three different energy regions.
It is well-known that the dynamical field degrees of freedom in the energy
region above CSSB scale are current quarks and gluons, and in the energy
region blow QC scale are pseudoscalar meson fields purely. In the energy
region between CSSB scale and QC scale, the CSSB implies that the
dynamical field degrees of freedom should be constituent 
quarks(quasi-particle of quarks), gluons and Goldstone bosons(massless
pseudoscalar meson fields) associated with CSSB. Nevertheless, it is still
debated whether pseudoscalar mesons are double counted when they are 
regarded as fundamental fields, as well as light bound states of
quarks\cite{MG84}. Theoretically, it is clear that there is not double
counting problem, since fundamental pseudoscalar mesons are light bound
states of light current quarks instead of heavy constituent quarks. 
Practically, however, due to lack of essential evidences to confirm this
point, the pseudoscalar mesons are still regared as composited fields of
constituent quarks in many works of low energy QCD in the
literatures\cite{NJL,ERT90,Chan85}. The purpose of this present
paper is to provide an evidence to confirm that pseudoscalar
mesons are fundamental dynamical degrees of freedom in the energy region
below CSSB scale indeed.  

Our analysis is based on to solve a puzzle which also lies low energy
strong interaction. In QCD below CSSB scale, the axial weak current with
light flavors can be defined with a non-unit coupling constant,
\begin{eqnarray*}
j_{5\mu}^a=g_{_A}\bar{\psi}\gamma_\mu\gamma_5\frac{\tau^a}{2}\psi,
\hspace{1in}(a=1,2,3),
\end{eqnarray*} 
where $\bar{\psi}=(\bar{u},\bar{d})$ and $\tau^a$ are Pauli
matrices. If we recognize that the wavefunction of nucleon has $SU(3)_{\rm
color}\times SU(2)_{\rm flavor}$ symmetry, the beta decay of neutron
requires $g_{_A}=0.75$\cite{MG84}. Meanwhile, the PCAC(partially conserved
axial current) hypothesis and Adler-Bell-Jackiw anomaly\cite{ABJ69} imply
axial current divergence
\begin{eqnarray}\label{1}
\pa^\mu j_{5\mu}^{(3)}={f_\pi\over 2} m_\pi^2\pi^0+\frac{N_c}{24\pi}g_{_A}
\alpha_{\rm e.m.}\ep^{\mu\nu\alpha\beta}F_{\mu\nu} F_{\alpha\beta},
\end{eqnarray}
where $f_\pi=185.2$MeV is the pion decay constant and $F_{\mu\nu}=\pa_\mu
A_\nu-\pa_\nu A_\mu$ is photon field strength. Thus in chiral limit, 
the transition matrix element for the $\pi^0\rightarrow \gamma\gamma$
reduces to
\begin{eqnarray}\label{2}
<\gamma(\ep_1,k_1),\gamma(\ep_2,k_2)|\pi^0(q)>
=(2\pi)^4\delta^4(q-k_1-k_2)(-i)\frac{2N_c}{3\pi f_\pi}
 g_{_A}\alpha_{\rm e.m.}\ep^{\mu\nu\alpha\beta}
 \ep_{1\mu}k_{1\nu}\ep_{2\alpha}k_{2\beta}.  
\end{eqnarray} 
The experimental data for $\Gamma(\pi^0\rightarrow\gamma\gamma)$ yields
axial coupling constant $g_{_A}=1$ for $N_c=3$. This value is very
different from one determined by $\beta$ decay of neutron. Then how to
resolve this puzzle is a open question. In this paper, we will provide a
possible explanation on this problem in the framework of constituent quark
model. In this interpretation, the psudoscalar mesons must be regarded as
independent dynamical field degrees of freedom in this energy region. The
renormalization of pseudoscalar-constituent quarks loop effects will lead
to aixal coupling constant $g_{_A}$ running with renormalization
scale. Our results for one-loop renormalization are that taking
$g_{_A}(\mu=m_N)=0.75$ as input, we predict $g_{_A}(\mu=m_\pi)=0.96$
which matchs with requirement of $\pi^0\rightarrow\gamma\gamma$ decay
well. This result is an evdience that pseudoscalar mesons are
fundamental fields in energy scale below CSSB. In order to confirm this
conclusion further, we will calculate the anomal decays
$\rho^\pm\rightarrow\gamma\pi^\pm$
and $\omega\rightarrow\gamma\pi^0$. In this type of decays, effects of
$g_{_A}$ are the leading order. Our calculations will be up to the next to
leading order of $N_c^{-1}$ expansion and including all order effects of
vector meson momentum expansion. The results strongly support prediction
of one-loop renormalization for axial coupling constant,
$g_{_A}(\mu=m_\rho)=0.77$.

The simplest version of chiral quark model was originated by
Weinberg\cite{Wein79}, and developed by Manohar and Georgi\cite{MG84}
provides a QCD-inspired description on the constituent quark model. 
At chiral limit, it is parameterized by the following $SU(3)_{_V}$
invariant chiral constituent quark lagrangian
\begin{eqnarray}\label{3}
{\cal L}_{\chi}&=&i\bar{\psi}(\sla{\pa}+\sla{\Gamma}+
  g_{_A}{\slash\!\!\!\!\Delta}\gamma_5-)\psi-m\bar{\psi}\psi
   +\frac{F^2}{16}Tr_f\{\nabla_\mu U\nabla^\mu U^{\dag}\}.
\end{eqnarray}
Here $Tr_f$ denotes trace in SU(3) flavour space,
$\bar{\psi}=(\bar{u},\bar{d},\bar{s})$ are constituent quark fields.
The $\Delta_\mu$ and $\Gamma_\mu$ are defined as follows,
\begin{eqnarray}\label{5}
\Delta_\mu&=&\frac{1}{2}\{\xi^{\dag}(\pa_\mu-ir_\mu)\xi
          -\xi(\pa_\mu-il_\mu)\xi^{\dag}\}, \nonumber \\
\Gamma_\mu&=&\frac{1}{2}\{\xi^{\dag}(\pa_\mu-ir_\mu)\xi
          +\xi(\pa_\mu-il_\mu)\xi^{\dag}\},
\end{eqnarray}
and covariant derivative are defined as follows
\begin{eqnarray}\label{6}
\nabla_\mu U&=&\pa_\mu U-ir_\mu U+iUl_\mu=2\xi\Delta_\mu\xi,
  \nonumber \\
\nabla_\mu U^{\dag}&=&\pa_\mu U^{\dag}-il_\mu U^{\dag}+iU^{\dag}r_\mu
  =-2\xi^{\dag}\Delta\xi^{\dag},
\end{eqnarray}
where $l_\mu=v_\mu+a_\mu$ and $r_\mu=v_\mu-a_\mu$ are linear combinations
of external vector field $v_\mu$ and axial-vector field $a_\mu$, $\xi$
associates with non-linear realization of spontanoeusly broken global
chiral symmetry $G=SU(3)_L\times SU(3)_R$ introduced by
Weinberg\cite{Wein68},
\begin{equation}\label{7}
\xi(\Phi)\rightarrow
g_R\xi(\Phi)h^{\dag}(\Phi)=h(\Phi)\xi(\Phi)g_L^{\dag},\hspace{0.5in}
 g_L, g_R\in G,\;\;h(\Phi)\in H=SU(3)_{_V}.
\end{equation}
Explicit form of $\xi(\Phi)$ is usual taken
\begin{equation}\label{8}
\xi(\Phi)=\exp{\{i\lambda^a \Phi^a(x)/2\}},\hspace{1in}
U(\Phi)=\xi^2(\Phi),
\end{equation}
where the Goldstone boson $\Phi^a$ are treated as pseudoscalar meson
octet. The constituent quark fields transform as matter fields of
SU(3)$_{_V}$,
\begin{equation}\label{9}
  \psi\lraw h(\Phi)\psi, \hspace{1in} \bar{\psi}\lraw
\bar{\psi}h^{\dag}(\Phi).
\end{equation}
$\Delta_\mu$ is SU(3)$_{_V}$ invariant field gradients and $\Gamma_\mu$
transforms as field connection of SU(3)$_{_V}$
\begin{equation}\label{12}
\Delta_\mu\lraw h(\Phi)\Delta_\mu h^{\dag}(\Phi), \hspace{0.8in}
\Gamma_\mu\lraw h(\Phi)\Gamma_\mu h^{\dag}(\Phi)+h(\Phi)\pa_\mu
  h^{\dag}(\Phi).
\end{equation}
Thus the lagrangian(~\ref{3}) is invariant under $G_{\rm global}\times
G_{\rm local}$.

Let us consider one-loop effects renormalization of lagrangian~(\ref{3})
\footnote{It should be pointed out that, this model as low energy
effective model of QCD is not completely renormalizable. The reason is
that more and more new divergent terms will apear when loop effects are
included. Fortunately, the terms in lagrangian~(\ref{3}) are still
renormalizable.}.   
In terms of defining the following ``renormalization'' quantities,
\begin{equation}\label{13}
\begin{array}{c}
\psi\rightarrow\psi_{_R}=Z_{\psi}^{-1/2}\psi,\hspace{1in}
g_{_A}\rightarrow g_{_{RA}}=Z_{g}g_{_A}, \\
m\rightarrow m_{_R}=m+\delta m,\hspace{1in}
F\rightarrow F_{_R}=Z_{_F}^{1/2}F, \\
\xi(\Phi)\rightarrow\xi_{_R}(\Phi)=Z_{\xi}^{-1/2}\xi(\Phi)\;\;\;
\Longrightarrow\;\;\;
\left\{\begin{array}{c}
  \Gamma_{\mu R}=Z_\xi^{-1}\Gamma_\mu, \\
     \Delta_{\mu R}=Z_\xi^{-1}\Delta_\mu, \\ U_{_R}=Z_\xi^{-1}U,
\end{array}\right.
\end{array} 
\end{equation}
the divergences yielded by loop-effects of constituent quarks and
massless Goldstone fields can be cancelled by divergent constants
$Z_{\psi},\;Z_{g},\;\delta m,\;Z_{_F}$ and $Z_{\xi}$. In this paper, we
focus our attention on renormalization of axial constant $g_{_A}$. The
diagrams in fig. 1 concern our purpose. In $\overline{\rm MS}$ scheme,
explicit calculation gives
\begin{eqnarray}\label{14}
Z_{\psi}&=&1-\frac{2m_{_R}^2}{\Lambda_{\chi}^2}g_{_{RA}}^2N_{\ep},
\hspace{1in} 
Z_{\psi}\delta m=-\frac{2m_{_R}^2}{3\Lambda_{\chi}^2}g_{_{RA}}^2N_{\ep}, 
\nonumber \\
Z_{\psi}Z_{\xi}&=&1-\frac{m_{_R}^2}{4\Lambda_{\chi}^2}g_{_{RA}}^2N_{\ep},
\hspace{1in}
Z_{\psi}Z_{\xi}Z_{g}^{-1}=1-\frac{m_{_R}^2}{\Lambda_{\chi}^2}(\frac{3}{2}
  +\frac{5}{12}g_{_{RA}}^2)N_{\ep},
\end{eqnarray}
where $\Lambda_{\chi}=2\pi f_\pi\simeq 1.2$GeV is CSSB scale and
\begin{eqnarray}\label{15}
N_{\ep}=\lim_{\ep\rightarrow 0}\frac{2}{\ep}-\gamma_{_E}+\ln{4\pi}.
\end{eqnarray}

\begin{figure}  
   \centerline{
   \psfig{figure=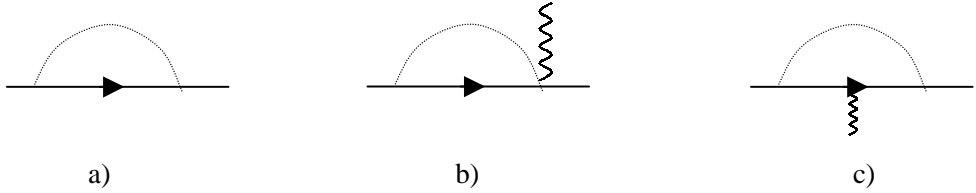,width=5.5in}}
\begin{center}
\begin{minipage}{5in}
   \caption{One-loop renormalization of lagrangian~(\ref{3}). The solid
line is constituent quarks, dot line denotes massless pseudoscalar meson
fields and wave line denotes external vector or axial-vector fields.} 
\end{minipage}
\end{center}
\end{figure}

Then renomalization lagrangian is written as follow
\begin{eqnarray}\label{16}
{\cal L}_{_R}&=&i\bar{\psi}_{_R}(\sla{\pa}+\sla{\Gamma_{_R}})
  \psi_{_R}-\{1-\frac{2m_{_R}^2}{3\Lambda_{\chi}^2}g_{_{RA}}^2
  \ln{\frac{\mu^2}{m^2}}+O(\frac{m_{_R}^4}{\Lambda_{\chi}^4})\}m_{_R}
  \bar{\psi}_{_R}\psi_{_R} \nonumber \\
&&+\{1+\frac{m_{_R}^2}{\Lambda_{\chi}^2}({3\over 2}-{4\over 3}g_{_{RA}}^2)
   \ln{\frac{\mu^2}{m^2}}+O(\frac{m_{_R}^4}{\Lambda_{\chi}^4})\}
   g_{_{RA}}\bar{\psi}_{_R}i\slash\!\!\!\!\Delta\gamma_5\psi_{_R}
+\frac{F_{_R}^2}{16}Tr_f\{\na_\mu U_{_R}\na^\mu U_{_R}^{\dag}\},
\end{eqnarray}
where $\mu$ is renomalization scale. The $\mu$-independence of
renormalization requires the constituent quark mass $m_{_R}$ and axial
constant $g_{_{RA}}$ should be $\mu$-dependent, and should satisfy(for
sake of convenience, we omit the subscript ``$R$'' in $m_{_R}$ and
$g_{_{RA}}$ hereafter) 
\begin{eqnarray}\label{17}
\frac{m(\mu^{\pr})}{m(\mu)}&=&1-\frac{2\bar{m}^2}{3\Lambda_{\chi}^2}
  \bar{g}_{_A}^2\ln{\frac{\mu^2}{\mu^{\pr 2}}}
  +O(\frac{m^4}{\Lambda_{\chi}^4}),\nonumber \\
\frac{g_{_A}(\mu^{\pr})}{g_{_A}(\mu)}&=&1+\frac{\bar{m}^2}
   {\Lambda_{\chi}^2}({3\over 2}-{4\over 3}\bar{g}_{_A}^2)
   \ln{\frac{\mu^2}{\mu^{\pr 2}}}+O(\frac{m^4}{\Lambda_{\chi}^4}),
\end{eqnarray}
where
\begin{eqnarray}\label{18}
\bar{m}=\frac{m(\mu)+m(\mu^{\pr})}{2},\hspace{1in}
\bar{g}_{_A}=\frac{g_{_A}(\mu)+g_{_A}(\mu^{\pr})}{2}.
\end{eqnarray}
Then inputting $m(\mu=m_\rho)=480$MeV
\footnote{The $O(p^4)$ chiral coupling constant
$L_5=(\frac{3}{8}g^2-\frac{N_c}{16\pi^2})\frac{m}{2B_0}$ is predicted in
this model. Here $g$ and $B_0$ are two constants absorbing divergence from
quark loops. In particular, $g=\pi^{-1}$ and $B_0=m_\pi^2/(m_u+m_d)\simeq
2$GeV is fitted at $\mu=m_\rho$\cite{Rho}. Thus $m=480$MeV is fitted by
$L_5(\mu=m_\rho)=(1.4\pm 0.5)\times 10^{-3}$.}  
and $g_{_A}(\mu=m_N)=0.75$, we have
\begin{eqnarray}\label{19}
m(\mu=m_N)&=&494{\rm MeV},\hspace{1in}m(\mu=m_\pi)=368{\rm MeV}
 \nonumber \\
g_{_A}(\mu=m_\rho)&=&0.77,\hspace{1.3in}g_{_A}(\mu=m_\pi)=0.96.
\end{eqnarray}
In particular, $g_{_A}(\mu=m_\pi)=0.96$ is agree with
$\pi^0\rightarrow\gamma\gamma$ requirement. 

Several remarks are necessary here. 1) The renormalization running of $m$
and $g_{_A}$ is rather larger. This point can be easily understood, since
there is no a rather small parameter to suppress loop effects in this
energy region. This is a feature of low energy QCD. For example, a low
energy coupling constant of chiral perturbative theory at $O(p^4)$, $L_5$,
is equal to $(2.2\pm 0.5)\times 10^{-3}$ at energy scale $\mu=m_\eta$ but
$(1.4\pm 0.5)\times 10^{-3}$ at energy scale $\mu=m_\rho$.
2) If pseudoscalar mesons are not independent dynamical field degrees of
freedom, all diagrams in fig.1 will be absent. Then we can not understand
why the value axial constant $g_{_A}$ should be equal to 0.75 in $\beta$
decay of neutron but be equal to 1 in $\pi^0\rightarrow\gamma\gamma$.
Therefore, our results in eqs.~(\ref{17}) and (\ref{16}) is a direct
support that massless pseudoscalar mesons as Goldstone bosons associating
CSSB are fundamental dynamical degrees of freedom. 3) The loop effects
also generate some high derivative terms contributing to axial constant,
e.g., $\bar{\psi}\pa^2\slash\!\!\!\!\Delta\gamma_5\psi$. However, we
expect that this type terms are suppressed by $p^2/\Lambda_{\chi}^2$
expansion(where $p$ denote four-momentum of axial current). For $\beta$
decay of neutron and $\pi^0\rightarrow\gamma\gamma$, the transition
momentum are both very small comparing with $\Lambda_\chi$. So the
contributions from high derivative terms are very small that it can not
explain the difference of axial constant in two reactions. 4) If we
inserted Adler-Bell-Jackiw anomal term with unit axial constant into
original lagrangian~(\ref{3}), unambiguously, it should cause double
counting, since triangle diagrams of constituent quarks also generate
Adler-Bell-Jackiw anomally. 5) In lagrangian~(\ref{3}) pseudoscalar mesons
are massless. It is consistent with Goldstone theorem for CSSB. 

In the following, for checking the result of eq.~(\ref{17}), we will
provide a complete calculation on decays
$\rho^\pm\rightarrow\gamma\pi^\pm$ and $\omega\rightarrow\gamma\pi^0$.
Since the chiral expansion converges slowly at energy scale of vector
meson masses\cite{Rho}, we have to include high order contribution of
chiral expansion into our results. In this paper, our calculation will be
up to the next to leading order of $N_c^{-1}$ expansion, and including all
important effects of momentum expansion.   

Due to WCCWZ realization for vector meson resonances\cite{Wein68,CCWZ69},
the lagrangian can be easily extended to included the lowest vector meson
resonances
\begin{eqnarray}\label{301}
{\cal L}_{\chi}&=&i\bar{\psi}(\sla{\pa}+\sla{\Gamma}+
  g_{_A}{\slash\!\!\!\!\Delta}\gamma_5-i\sla{V})\psi-m\bar{\psi}\psi    
   +\frac{F^2}{16}Tr_f\{\nabla_\mu U\nabla^\mu U^{\dag}\}
   +\frac{1}{4}m_0^2Tr_f\{V_\mu V^{\mu}\}.
\end{eqnarray}
Here $V_\mu$ denotes vector meson octet and singlet, or more convenience,
due to OZI rule, they are combined into a singlet ``nonet'' matrix
\begin{equation}\label{4}
   V_\mu(x)={\bf \lambda \cdot V}_\mu =\sqrt{2}
\left(\begin{array}{ccc}
       \frac{\rho^0_\mu}{\sqrt{2}}+\frac{\omega_\mu}{\sqrt{2}}
            &\rho^+_\mu &K^{*+}_\mu   \\
    \rho^-_\mu&-\frac{\rho^0_\mu}{\sqrt{2}}+\frac{\omega_\mu}{\sqrt{2}}
            &K^{*0}_\mu   \\
       K^{*-}_\mu&\bar{K}^{*0}_\mu&\phi_\mu
       \end{array} \right).
\end{equation}
It transforms homogeneously under SU(3)$_{_V}$
\begin{equation}\label{10}
 V_\mu\lraw h(\Phi)V_\mu h^{\dag}(\Phi),
\end{equation}
Thus the lagrangian~(\ref{301}) is still invariant under $G_{\rm
global}\times G_{\rm local}$.
The effective action describing meson interaction can be obtained via
integrating over degrees of freedom of fermions in lagrangian~(\ref{301})
\begin{equation}\label{20}
e^{iS_{\rm eff}}\equiv\int{\cal D}\bar{q}{\cal D}qe^{i\int d^4x{\cal
   L}_\chi(x)}=<vac,out|in,vac>_{V,\Delta,\Gamma},
\end{equation}
where $<vac,out|in,vac>_{V,\Delta,\Gamma}$ is vacuum expectation value
in presence external sources. The above path integral can be performed
explicitly, and heat kernal method\cite{Sch51,Ball89} has been used to
regulate the result. However, this method is extremely difficult to
compute very high order contributions in practice. This diffculty can be
overcomed via calculating one-loop diagrams of constituent quarks
directly. This method can capture all high order contributions of the
chiral expansion.

In interaction picture, the equation(~\ref{20}) is rewritten as follow  
\begin{eqnarray}\label{21}
e^{iS_{\rm eff}}&=&<0|{\cal T}_qe^{i\int d^4x{\cal L}^{\rm I}_\chi(x)}|0>
       \nonumber \\
 &=&\sum_{n=1}^\infty i\int d^4p_1\frac{d^4p_2}{(2\pi)^4}
  \cdots\frac{d^4p_n}{(2\pi)^4}\tilde{\Pi}_n(p_1,\cdots,p_n)
  \delta^4(p_1-p_2-\cdots-p_n) \nonumber \\
&\equiv&i\Pi_1(0)+\sum_{n=2}^\infty i\int \frac{d^4p_1}{(2\pi)^4}   
  \cdots\frac{d^4p_{n-1}}{(2\pi)^4}\Pi_n(p_1,\cdots,p_{n-1}),
\end{eqnarray}
where ${\cal T}_q$ is time-order product of constituent quark fields,    
${\cal L}_{\chi}^{\rm I}$ is interaction part of lagrangian(~\ref{301}),
$\tilde{\Pi}_n(p_1,\cdots,p_n)$ is one-loop effects of constituent quarks
with $n$ external sources, $p_1,p_2,\cdots,p_n$ are four-momentas of $n$
external sources respectively and
\begin{equation}\label{22}
\Pi_n(p_1,\cdots,p_{n-1})=\int d^4p_n\tilde{\Pi}_n(p_1,\cdots,p_n)
  \delta^4(p_1-p_2-\cdots-p_n).
\end{equation}
To get rid of all disconnected diagrams, we have
\begin{eqnarray}\label{23}
S_{\rm eff}&=&\sum_{n=1}^\infty S_n, \nonumber \\
S_1&=&\Pi_1(0),  \\
S_n&=&\int \frac{d^4p_1}{(2\pi)^4}\cdots\frac{d^4p_{n-1}}
  {(2\pi)^4}\Pi_n(p_1,\cdots,p_{n-1}), \hspace{0.8in}(n\geq 2)\nonumber.
\end{eqnarray}
Hereafter we will call $S_n$ as $n$-point effective action.

At the leading order of $N_c^{-1}$ expansion, the 3-point $\rho\gamma\pi$
and $\omega\gamma\pi$ effective action is generated by triangle diagram of
constituent quarks
\begin{eqnarray}\label{24}
\Pi_3(q,k)&=&\frac{N_c}{3\pi^2 gf_\pi}eg_{_A}
  \int_0^1dx_1\int_0^{1-x_1}dx_2\{1-(x_1+x_2)(1-x_1-x_2)\frac{q^2}{2m^2}
  \}^{-1}\nonumber \\ &&\times\ep^{\mu\nu\alpha\beta}
  q_\alpha k_\beta\{\rho^i_\mu(q)A_\nu(k)\pi_i(q-k)
  +3\omega_\mu(q)A_\nu(k)\pi^0(q-k)\},
\end{eqnarray}
where $g$ is an universal coupling constant which absorbs
nonrenormalizable logarithmic divergence from constituent quark loops. In
the above equation, $g$ is from vector meson field normalization
\begin{eqnarray*}
\rho_\mu^i\rightarrow\frac{1}{g}\rho_\mu^i,\hspace{1in}
\omega_\mu\rightarrow\frac{1}{g}\omega_\mu.  
\end{eqnarray*}
In ref.\cite{Rho}, $g=\pi^{-1}$ has been fitted by the first KSRF sum
rule\cite{KSRF}.

It is rather complicate to calculate contribution from meson one-loop,
i.e., the next to leading order contribution of $N_c^{-1}$ expansion. The
diagrams in fig.2 concern to our calculation. Here it is requirement of
unitarity of $S$-matrix to sum over chain-like approximation in
fig.2-b).

\begin{figure}
   \centerline{
   \psfig{figure=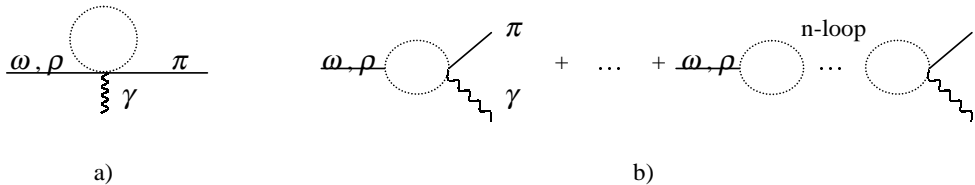,width=5.5in}}
\begin{center}
\begin{minipage}{5in}
   \caption{Meson loop correction to $\rho\rightarrow\gamma\pi$ and
$\omega\rightarrow\gamma\pi$ decay. The dot internal lines denote
physical pseudoscalar meson fields(massless pion and massive $K$ and
$\eta$. a) Tadpole loop. b) Chain-like approximation.}
\end{minipage}
\end{center}
\end{figure}

For obtain the tadpole correction in fig.2-a), at leading order of 
$N_c^{-1}$ expansion, vertices of 
$\rho\gamma\pi-KK(\eta\eta)$ and $\omega\gamma\pi-KK(\eta\eta)$ are needed
if we treat pion as massless particle. In scheme of dimensional
regularization, only the vertices generated by triangle diagram of
constituent quarks yield non-zero contribution,
\begin{eqnarray}\label{25}
\delta\Pi_3(q,k)&=&\frac{N_c}{16\pi^2 gf_\pi}eg_{_A}
  \int_0^1dx_1\int_0^{1-x_1}dx_2\{1-(x_1+x_2)(1-x_1-x_2)\frac{q^2}{2m^2}
  \}^{-1}\ep^{\mu\nu\alpha\beta}q_\alpha k_\beta \pi^l\vphi^a\vphi^b 
   \nonumber\\ &&
  Tr\{\frac{1}{3}(\rho_\mu^i(q)\lambda^i+3\omega_\mu(q)+eQA_\mu(q))
  (\rho_\nu^j(k)\lambda^j+3\omega_\nu(k)+eQA_\nu(k))
  ([\lambda^a, \lambda^l]\lambda^b+\lambda^a[\lambda^l,\lambda^b]) 
    \nonumber \\&&
  +eA_\mu(q)([\lambda^a,Q]\lambda^b+\lambda^a[Q,\lambda^b])
   (\rho_\mu^i(k)\lambda^i+3\omega_\mu(k))\pi^l\nonumber \\&&
  +e(\rho_\mu^i(q)\lambda^i+3\omega_\mu(q))A_\nu(k)
    ([\lambda^a,Q]\lambda^b+\lambda^a[Q,\lambda^b])\pi^l\},
    \hspace{0.5in} (i,j,l=1,2,3;a,b=4,...,8)
\end{eqnarray}
where $Q={\rm diag}\{2/3,-1/3,-1/3\}$ is charge operator of quark fields,
$\vphi^a$ denotes $K-$ or $\eta$-meson fields in internal line. For sake
of convenience, we can assume that the masses of $K$ and $\eta$ are
degenerated. Then integral over $\vphi$ fields, we have tadpole correction
to $\rho\gamma\pi$ and $\omega\gamma\pi$ vertices
\begin{eqnarray}\label{26}
\Pi_3^{tad}(q,k)=-\frac{4}{3}\zeta\Pi_3(q,k),\hspace{1in}
\zeta=\frac{m_{_K}^2}{8\pi^2f_\pi^2}(\frac{4\pi\mu^2}{m_{_K}^2})^{\ep/2}
\Gamma(1-\frac{D}{2})\equiv \frac{m_{_K}^2}{8\pi^2f_\pi^2}\lambda,
\end{eqnarray}
where $\Pi_3(q,k)$ is defined in eq.(~\ref{24}), $\lambda$ absorb the
quadratic divengence from meson loops. $\lambda\simeq 2/3$ has been fitted
by Zweig rule\cite{Rho}.

\begin{figure}
   \centerline{
   \psfig{figure=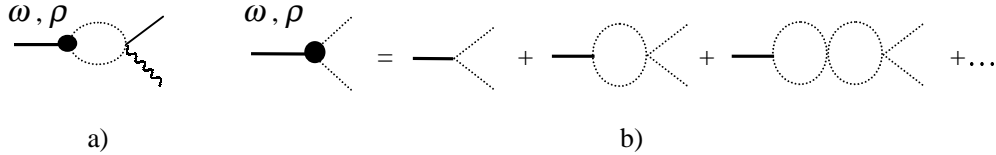,width=5.5in}}
\begin{center}
\begin{minipage}{5in}
\caption{The chain-like approximation in fig.2-b) is equivalent to
contribution of fig.3-a). It is generated through contracting an
effective VPP(where ``V'' denotes $\rho$ or $\omega$ meson, ``P''
denotes pseudoscalar meson) vertex and  $\gamma\pi$-PP vertex. Here the
effective VPP vertex includes not only tree level vertex but also meson
loop contribution(see fig.3-b).} 
\end{minipage}
\end{center}
\end{figure}
 
Next we will calculate the contribution in fig.2-b). It is equivalent to
calculate the diagram in fig.3-a). In other words, we need to an effective
VPP(where ``V'' denotes $\rho$ or $\omega$ meson, ``P'' denotes
pseudoscalar meson octet) vertex and $\gamma\pi$-PP vertex. In fig.3-b) we
have shown that the effective VPP vertex includes not only tree level
vertex but also meson loop contribution. 

At leading order of $N_c^{-1}$ expansion, the anomal $\gamma\pi$-PP vertex
is generated by both of triangle and box diagrams of constituent quarks,
\begin{eqnarray}\label{27}
&&\Gamma_{\gamma\rightarrow 3P}(q_2,k_1,k_2,k_3)=\frac{3i}{8f_\pi^2}
 eB(q^2)\ep^{\mu\nu\alpha\beta}A_\mu(q_2)k_{1\nu}k_{2\alpha}k_{3\beta}
  Tr_f\{QP(k_1)P(k_2)P(k_3)\}, \nonumber \\
&&B(q^2)=\frac{N_c}{3\pi^2f_\pi}g_{_A}(\int_0^1dx_1\int_0^{1-x_1}dx_2
 \{1-\frac{q^2}{2m^2}(x_1+x_2)(1-x_1-x_2)\}^{-1}-\frac{g_{_A}^2}{6}),
\end{eqnarray}
where $q^2=(k_1+k_2)^2$. The effective $\rho$-PP vertex and $\omega-KK$
vertex(here we omit those vertices suppressed by isospin conversation) 
have been derived in ref.\cite{Rho}
\begin{eqnarray}\label{28}
\Gamma_{\rho PP}(q,k_1,k_2)&=&\frac{1}{2}g_{\rho PP}(q^2)
  (q^2k_{2\mu}-q_\mu q\cdot k_2)Tr_f\{\rho^\mu(q)P(k_1)P(k_2)\},
     \nonumber \\
\Gamma_{\omega KK}(q,k_1,k_2)&=&\frac{1}{2}g_{\omega KK}(q^2)  
  (q^2k_{2\mu}-q_\mu q\cdot k_2)\omega^\mu
 \{(K^+(k_1)K^-(k_2)+K^0(k_1)\bar{K}^0(k_2))+(k_1\leftrightarrow k_2)\},
\end{eqnarray}
with
\begin{eqnarray}\label{29}
g_{\rho PP}(q^2)&=&\frac{A_1(q^2)+g_{_A}^2A_2(q^2)}
  {gf_\pi^2(1+2\zeta)(1+\Sigma(q^2))}, \nonumber \\
g_{\omega KK}(q^2)&=&\frac{A_1(q^2)+g_{_A}^2A_2(q^2)}
  {gf_\pi^2(1+2\zeta)(1-2\Sigma_K(q^2))},
\end{eqnarray}
where
\begin{eqnarray}\label{30}
A_1(q^2)&=&g^2-\frac{N_c}{\pi^2}\int_0^1dx\cdot x(1-x)
  \ln{(1-\frac{x(1-x)q^2}{m^2})},\nonumber \\
A_2(q^2)&=&-g^2+\frac{N_c}{2\pi^2}\int_0^1dx_1\int_0^1dx_2
  (x_1-x_1^2x_2)\{1+\frac{m^2}{m^2-x_1(1-x_1)(1-x_2)q^2}\nonumber \\
 &&\hspace{1in}+\ln{(1-\frac{x_1(1-x_1)(1-x_2)q^2}{m^2})}\}, 
  \nonumber \\
\Sigma(q^2)&=&\{1+\frac{q^2(A_1(q^2)+2g_{_A}^2A_2(q^2))}
 {2f_\pi^2(1+11\zeta/3)}\}(4\Sigma_\pi(q^2)-2\Sigma_K(q^2)),\\
\Sigma_\pi(q^2)&=&\frac{q^2}{16\pi^2f_\pi^2}\{\frac{\lambda}{6}
  +\int_0^1dt\cdot t(1-t)\ln{\frac{t(1-t)q^2}{m_{_K}^2}}
  +\frac{i}{6}Arg(-1)\theta(q^2-4m_\pi^2)\},\nonumber \\
\Sigma_{K}(q^2)&=&\frac{1}{16\pi^2f_\pi^2}\{\lambda(m_{_K}^2-\frac{q^2}{6})
  +\int_0^1dt[m_{_K}^2-t(1-t)q^2]\ln{(1-\frac{t(1-t)q^2}{m_{_K}^2})}\}.
   \nonumber
\end{eqnarray}

Then due to soft-pion theorem, the pseudoscalar meson loops in fig.2-b)
contribution to $\rho\gamma\pi$ and $\omega\gamma\pi$ vertices as follow
\begin{eqnarray}\label{31}
\Pi_3^{1-loop}(q,k)&=&-eq^2g_{\rho PP}(q^2)[A(q^2)+2A(0)][\Sigma_\pi(q^2)
  -\frac{1}{2}\Sigma_K(q^2)]\ep^{\mu\nu\alpha\beta}q_\alpha k_\beta
  \rho_\mu^i(q)A_\nu(k)\pi_i(q-k)\nonumber \\
&&+\frac{1}{2}eq^2g_{\omega KK}(q^2)[A(q^2)+2A(0)]
  \Sigma_K(q^2)\ep^{\mu\nu\alpha\beta}q_\alpha k_\beta
  \omega_\mu(q)A_\nu(k)\pi^0(q-k).
\end{eqnarray}
Here $\rho\gamma\pi$ coupling receives contributions from both of 
pion-loop and $K$-loop, but $\omega\gamma\pi$ coupling receives dominant
contribution from $K$-loop only.

Eq.(\ref{31}) tegother with eqs.(\ref{24}) and (\ref{26}) give
``complete'' $\rho\gamma\pi$ and $\omega\gamma\pi$ coupling up to the next
to leading of $N_c^{-1}$ expansion at least,
\begin{eqnarray}\label{32}
\Pi_3^{c}(q,k)&=&e\ep^{\mu\nu\alpha\beta}q_\alpha
  k_\beta\{g_{\rho\gamma\pi}(q^2)\rho_\mu^i(q)A_\nu(k)\pi_i(q-k)
+g_{\omega\gamma\pi}(q^2)\omega_\mu(q)A_\nu(k)\pi^0(q-k)\},\nonumber \\
g_{\rho\gamma\pi}(q^2)&=&\frac{N_c}{3\pi^2gf_\pi}g_{_A}(1-\frac{4}{3}
 \zeta)\int_0^1dx_1\int_0^{1-x_1}dx_2\{1-(x_1+x_2)(1-x_1-x_2)
 \frac{q^2}{2m^2}\}^{-1} \nonumber \\
 &&-q^2g_{\rho PP}(q^2)[A(q^2)+2A(0)]
 [\Sigma_\pi(q^2)-\frac{1}{2}\Sigma_K(q^2)], \\
g_{\omega\gamma\pi}(q^2)&=&\frac{N_c}{\pi^2gf_\pi}g_{_A}(1-\frac{4}{3}   
 \zeta)\int_0^1dx_1\int_0^{1-x_1}dx_2\{1-(x_1+x_2)(1-x_1-x_2)
 \frac{q^2}{2m^2}\}^{-1} \nonumber \\
 &&+\frac{1}{2}q^2g_{\omega KK}(q^2)[A(q^2)+2A(0)]\Sigma_K(q^2).
\end{eqnarray}
For $g_{_A}(\mu=m_\rho)=0.77$, the above results yield
\begin{eqnarray}\label{33}
B(\rho^\pm\rightarrow\pi^\pm\gamma)=4.88\times 10^{-4},
\hspace{1in}B(\omega\rightarrow\pi^0\gamma)=8.9\%.
\end{eqnarray}
These results agree with data\cite{PDG98} 
$B(\rho^\pm\rightarrow\pi^\pm\gamma)=(4.5\pm 0.5)\times 10^{-4}$ and
$B(\omega\rightarrow\pi^0\gamma)=(8.5\pm 0.5)\%$ well. If we take
$g_{_A}=1$ as a comparison, the theoretical predictions are
\begin{eqnarray}\label{34}
B(\rho^\pm\rightarrow\pi^\pm\gamma)=8.33\times 10^{-4},
\hspace{1in}B(\omega\rightarrow\pi^0\gamma)=15.0\%.
\end{eqnarray}
These results obviously disagree with data. Therefore, the theoretical
prediction~(\ref{33}) provides an evidence to confirm the result in
eq.~(\ref{17}).

To conclude, we discuss the one-loop renormalization in Manohar-Georgi
model. This renormalization leads to running of axial coupling constant
and constituent quark mass. Inputting $g_{_A}(\mu=m_N)=0.75$ which agree
with $\beta$ decay of neutron, the prediction $g_{_A}(\mu=m_\pi)=0.96$
agree with reqiurement of $\pi^0\rightarrow\gamma\gamma$ decay. We also
notice that, if massless pseudoscalar meson fields were not independent
degrees of freedom in energy between CSSB scale and QC scale, we could not
interpret $\beta$ decay of neutron and $\pi^0\rightarrow\gamma\gamma$
decay simultaneously. For example, because axial constant is not unit in
Extend Nambu-Jona-Lasinio model\cite{NJL}, the model can not yield right
prediction for $\pi^0\rightarrow\gamma\gamma$ decay. The similar problem
also exists in the models of ref.\cite{Chan85}. Therefore, our result
in this paper can be reagrd as an evidence that pseudoscalar meson fields
are independent degrees of freedom in energy between CSSB scale and QC
scale.

The renormalization of axial constant also predicts
$g_{_A}(\mu=m_\rho)=0.77$. For confirming this prediction, we calculate
anomal vector meson decays, $\rho^\pm\rightarrow\gamma\pi$ and
$\omega\rightarrow\gamma\pi$. In this type of decays, effects of $g_{_A}$
is leading order. Our results are up to the next to leading order of
$N_c^{-1}$ expansion and include all order effects of vector meson
momentum expansion. The theoretical predictions for branch ratios of these
two decays agree with results of renormalization and is against
$g_{_A}=1$.

\end{document}